\documentclass[11pt,preprint]{aastex}
\usepackage{epsfig}
\usepackage{amsmath}
\usepackage{multirow}

\begin{document}
\title{Diffuse PeV neutrinos from gamma-ray bursts}
\author{Ruo-Yu Liu\altaffilmark{1,2,3} and Xiang-Yu Wang\altaffilmark{1,4}}
\altaffiltext{1}{School of Astronomy and Space Science, Nanjing University,
Nanjing, 210093, China} \altaffiltext{2}{Max-Planck-Institut f\"ur Kernphysik, 69117 Heidelberg, Germany
}\altaffiltext{3}{Fellow of the International Max Planck Research School for Astronomy and Cosmic Physics at the University of Heidelberg (IMPRS-HD)}\altaffiltext{4}{Key laboratory of Modern
Astronomy and Astrophysics (Nanjing University), Ministry of
Education, Nanjing 210093, China}

\begin{abstract}
The IceCube collaboration  recently reported the potential detection
of two cascade neutrino events in the energy range 1-10 PeV. We
study the possibility that these PeV neutrinos are produced by
gamma-ray bursts (GRBs), paying special attention to the
contribution by untriggered GRBs that elude detection due to their
low photon flux. Based on the luminosity function, rate distribution
with redshift and spectral properties of GRBs, we generate,  using
Monte-Carlo simulation, a GRB sample  that reproduce the observed
fluence distribution of Fermi/GBM GRBs and an accompanying sample of
untriggered GRBs simultaneously. The neutrino flux of every
individual  GRBs is calculated in the standard internal shock
scenario, so that the accumulative flux of the whole samples can be
obtained. We find that the neutrino flux in PeV energies produced by
untriggered GRBs is about 2 times higher than that produced  by the
triggered ones. Considering the  existing  IceCube limit on the
neutrino flux of triggered GRBs, we find that the total flux of
triggered and untriggered GRBs can reach at most a level of $\sim
10^{-9}\, \rm GeVcm^{-2}s^{-1}sr^{-1}$, which is insufficient to
account for the reported two  PeV neutrinos. Possible contributions
to diffuse neutrinos by low-luminosity GRBs and  the earliest
population of GRBs are also discussed.

\end{abstract}
\keywords{ diffuse background : neutrinos --- gamma rays: bursts }

\section{Introduction}
Detection of high-energy neutrinos would be an important step
towards identifying the origin of cosmic rays. Interestingly,
through analysis designed to search for ultra high-energy ($\ga$
PeV) and high quality shower events over a period of 672.7 live days
between 2010 and 2012 with IceCube, two PeV neutrino-induced cascade
events have been found  \citep{Aya12}.  The two events were observed
on August 9, 2011 and January 3, 2012, and each have an energy of
approximately 1-10 PeV. The observation of two events with an
expected background of 0.14 events from the atmospheric neutrinos
corresponds to a significance of $2.36\sigma$. Given this modest
statistical significance, we cannot at this time be confident that
they are of astrophysical origin.

Gamma-ray bursts (GRBs)  have been proposed as a potential source of
ultra--high energy cosmic rays (UHECRs) \citep[e.g.][]{Waxman95,
Vietri95, Dermer02}, and high energy neutrino emission has been
predicted to be produced in the dissipative fireballs, where cosmic
ray protons interact with fireball photons \citep[e.g.][]{Waxman97,
Dermer06, Guetta04, Murase06b, Wang09}. Gamma-ray bursts are an
attractive possibility for the reported two PeV neutrinos because
the predicted neutrino emission in the standard internal shock model
peaks at PeV energies. Notably, previous searches of GRB neutrinos
with IceCube via stacking analysis of many triggered GRBs have yield
none detection\citep{IC10, IC11a, IC12}. The approach to search for
neutrinos from triggered GRBs is through analysis of upgoing
neutrinos that are correlated in time and direction with known GRBs,
which is different from that of detecting diffuse neutrinos. The
most stringent constraint is given by the combined analysis of the
IceCube 40- and 59-string (IC40+59) data of upgoing neutrinos from
215 triggered  GRBs, which leads to an upper limit that is claimed
to be below the   expected flux\citep{IC12}{\footnote{This claim is
based on the theoretic flux obtained  with the formula
in\citet{Guetta04}. Recent calculations show, however, that the
previous calculation may overestimate the flux of GRB
neutrinos\citep{Li12, Huemmer12, He12}. }}.  Beside the contribution
by triggered GRBs for which the stacking analysis has been
performed, GRBs that do not trigger detectors due to their low flux
may also contribute to the diffuse neutrino flux. The purpose of
this paper is to study whether the sum contribution of triggered
GRBs and untriggered GRBs can explain the reported two PeV neutrinos
under the constraint that the stacked neutrino flux of 215 triggered
GRBs is below the IC40+59 sensitivity.  {The untriggered GRBs
defined in our paper are normal GRBs (i.e. belonging to the normal
GRB population) that do not trigger the current detectors. They are
normal GRBs in the low-luminosity end of the luminosity function
(for normal GRBs) and/or locating at high redshifts and hence they
can elude the detection by current GRB detectors (e.g. Fermi/GBM)
due to the limited sensitivity.}

Previous estimates of diffuse neutrino flux from GRBs are made
through integrating the neutrino flux over the whole luminosity
range and the redshift range of GRBs, assuming that the their number
density follows a given luminosity function and  redshift
distribution function \citep[e.g.][]{Gupta07, He12, Cholis12}. The
estimates show that the total flux from all the GRBs (including both
the triggered ones and the untriggered ones) reach a level of $\sim
10^{-9}-10^{-8}\rm GeVcm^{-2}s^{-1}sr^{-1}$\citep[]{Gupta07, He12,
Cholis12}, which depends on the luminosity function and redshift
distribution that are used. These estimates made  simplified
assumptions about some properties of GRBs, e.g. assuming the same
photon spectrum  for every GRBs, and do not consider the existing
IceCube limit on triggered GRBs. To overcome this simplification, we
here present a new approach to calculate the diffuse neutrino flux.
We generate a GRB sample with Monte-Carlo simulation {\footnote { A
similar approach of sampling the bursts individually using a
Monte-Carlo algorithm was made by \citet{Baerwald12a}  to obtain a
more complete sample of GRBs. }}, requiring that the triggered
population of GRBs reproduce the observed properties of Fermi/GBM
GRBs. A population of dim, untriggered GRBs will be generated
simultaneously in the simulation. Then, the neutrino flux of every
individual  GRBs are calculated, so that the accumulative flux of
the whole sample can be obtained.

The rest part of this paper is arranged as follows. In \S 2, we
describe the properties of GRBs and some empirical relations used in
our simulation generating the GRB sample. We calculate the
accumulative neutrino flux from both triggered GRBs and untriggered
GRBs in the generated sample and compare them with the reported two
PeV events in \S 3. We give our discussion and conclusion in \S 4
and 5. Throughout the paper, we use eV for particle energy, c.g.s
units for other quantities and denote by $Q_x$ the value of the
quantity $Q$ in units of $10^x$ unless specified.

\section{Simulation of a GRB sample}
The distributions of GRBs in luminosity and in redshift are assumed
to be independent  of each other in this work, i.e.,
$n(L,z)=\phi(L)\rho(z)$, where $\phi(L)$ and $\rho(z)$ are
luminosity function and redshift distribution function of GRBs
respectively. \citet[][hereafter, W10]{Wanderman10} obtained the
luminosity function and cosmic rate by inverting directly the
redshift - luminosity distribution of observed long Swift GRBs,
which are, respectively, given  by
\begin{equation}\label{LF}
\phi(L)=\left\{
\begin{array}{ll}
\left(\frac{L}{L_*}\right)^{-a_1}, & L < L_*\\
\left(\frac{L}{L_*}\right)^{-a_2}, & L \geq L_*
\end{array}
\right.
\end{equation}
and
\begin{equation}\label{SF}
\rho(z)=\rho_0{\rm Gpc^{-3}yr^{-1}}\left\{
\begin{array}{ll}
(1+z)^{n_1}, & z < z_*\\
(1+z)^{n_2}(1+z_*)^{n_1-n_2}, & z \geq z_*
\end{array}
\right.
\end{equation}
where $a_1=1.2$, $a_2=2.4$, $L_*=10^{52.5}$ergs$^{-1}$, $n_1=2.1$,
$n_2=-1.4$, $z_*=3.1$, and $\rho_0=1.3$Gpc$^{-3}$yr$^{-1}$ are  the
best fit parameters. Note that $L$ is the  isotropic, peak
bolometric luminosity of GRBs in energy range 1\,keV---10\,MeV. The
low- and high- luminosity ends of the luminosity function are set as
$10^{50}$ergs$^{-1}$ and $10^{54}$ergs$^{-1}$ respectively, and the
maximum redshift is set as $z_{\rm max}=8$. We generate a sample of
GRBs that follow the luminosity distribution and redshift
distribution given by Eq.~\ref{LF} and Eq.~\ref{SF} with Monte-Carlo
simulation. The isotropic energy of each GRBs is obtained through
the relation
\begin{equation}
{\rm log}\,E_{\rm iso,52}= 1.07{\rm log}\,L_{\rm iso,52}+(0.66\pm
0.54)
\end{equation}
which is obtained from the data in \citet{Ghirlanda12}. The  photon
spectra of GRBs can be described by an empirical function known as
the Band function \citep{Band93}, which can be written as
\begin{equation}
N(E)=N0\left\{
\begin{array}{ll}
\left(\frac{E}{E_0}\right)^{\alpha}e^{-E/E_0}, & E < (\alpha-\beta)E_0\\
\left(\frac{E}{E_0}\right)^{\beta}(\alpha-\beta)^{\alpha-\beta}e^{\beta-\alpha}, & E \geq (\alpha-\beta)E_0
\end{array}
\right.
\end{equation}
where $N_0$ is the normalization constant and $\alpha$, $\beta$ are
the photon indices for low-- and high--energy spectra respectively.
Generally, $\alpha>-2$ and $\beta<-2$, and the energy spectrum peaks
at $E_{\rm peak}=(2+\alpha)E_0$. The values of $\alpha$ and $\beta$
are generated following the observed distribution of BATSE long GRBs
in the simulation and only bursts with $\beta<-2$ are adopted
\citep{Kaneko06}.  The rest frame peak energy of GRBs is found to be
correlated with their isotropic energy and luminosity, which are
known as the Amati relation and the Yonetoku relation, respectively
\citep{Amati02, Yonetoku04}. In this work, to generate the peak
energies of GRBs in the simulation, we adopt the $E_{\rm
peak}-E_{\rm iso}$ relation recently found by \citet{Ghirlanda12},
i.e.
\begin{equation}
\begin{split}
{\rm log}\left(\frac{E_{\rm peak}}{\rm keV}\right)&={\rm log}\left(\frac{E_{\rm peak}^{\rm ob}(1+z)}{\rm keV}\right)\\
&=0.56{\rm log}\left(\frac{E_{\rm iso}}{\rm erg}\right)-26.06\pm 1.14,
\end{split}
\end{equation}
where $E_{\rm peak}^{\rm ob}$ is the observer frame peak energy
(hereafter the subscript "ob" represents  the quantity in the
observer frame).  Since the neutrino emission is produced dominantly
by long duration GRBs,   we only include bursts with $T_{90}
>2$s in our sample. Following \citet{Kakuwa12}, we determine
$T_{90}$ roughly by $T_{90}=(1+z)(E_{\rm iso}/L_{\rm ave})$ where
$L_{\rm ave}=0.3L_{\rm iso}$ is the average luminosity of a GRB.

To compare our simulated sample with observed GRBs by
\emph{Fermi}/GBM, the trigger threshold in our simulation is set to
be the same as that of \emph{Fermi}/GBM, i.e.,
0.74\,photons\,cm$^{-2}$s$^{-1}$\citep{Meegan09}. Taking into
account the effect of Earth occultations and the South Atlantic
Anomaly (SAA) passages, about 65\% GRBs above the \emph{Fermi}/GBM
trigger threshold can be observed. To reproduce the GBM detection
rate of $\sim 250\,$GRBs per year,  the value of $\rho_0$ is
required to be $1.01\rm \,Gpc^{-3}yr^{-1}$, which is within the
$1\sigma$ error of the best--fit value obtained by W10. For this
value of $\rho_0$, the total event rate of GRBs in the luminosity
range from $10^{50}{\rm erg s^{-1}}$ to $10^{54} {\rm erg s^{-1}}$
and redshift range from $z=0$ to $z=8$ is $\sim 8000$ GRBs per year
in the generated sample so the vast majority of them are untriggered
GRBs. {The total number of GRBs occurred from $z=0$ to $z=8$ per
year is obtained  through $N=\int 4\pi
D_c^2(z)\frac{\rho(z)}{1+z}dD_c$, where
$D_c(z)=\int_0^z\frac{cdz}{H(z)}$ with
$H(z)=H_0\sqrt{(1+z)^3\Omega_M+\Omega_\Lambda}$ being the Hubble
constant at redshift $z$ in a flat universe.}

In Fig.~1, we compare the 10--1000keV fluence distributions
$dP/d({\rm log}F_{\gamma})$  of  GRBs above the trigger threshold in
our simulated sample with that of 482 \emph{Fermi}/GBM
bursts\footnote{From GRB~080714A to GRB~100704, data are taken from
http://heasarc.gsfc.nasa.gov/W3Browse/all/fermigbrst.html}. The K-S
test of two distributions returns a high accepting probability of
89.5\%. The total gamma-ray fluence of GRBs in 10--1000keV above the
trigger threshold in our simulated sample is $0.0048\,{\rm erg
\,cm^{-2}}$ when the number of bursts is normalized to be 482, which
also agree well with the total gamma-ray fluence, $0.0044\,{\rm erg
\,cm^{-2}}$, of the observed 482 GBM bursts.

In  Fig.~2 we show the distribution of the cumulative bolometric
fluence $\sum_i F_{\gamma,i}(>F_{\gamma, th})$ above a certain flux
$F_{\gamma, th}$ for the subsample of both triggered and untriggered
GRBs. {As is shown, the total gamma ray fluence of the untriggered
GRBs ($F_{\gamma,\rm unt}=0.0136\, \rm erg\,cm^{-2}$) is 2-3 times
larger than that of the triggered ones ($F_{\gamma,\rm
tri}=0.0053\,\rm erg\,cm^{-2}$). This is mainly due to two aspects.
First,  the number of untriggered GRBs is much larger than that of
triggered GRBs, although the luminosity of untriggered GRBs is on
average much lower than that of triggered GRBs; Second, a fraction
(about 35\%) of bright GRBs occulted by the Earth also contribute to
the gamma-ray fluence in the untriggered GRBs.} This demonstrates
that the total contribution of untriggered GRBs can be important.

There is significant variation in the luminosity function and the
redshift distribution of GRBs in the literature, see e. g.
\citet[][hereafter, L07]{Liang07}{\footnote{The luminosity function
obtained by  L07 is
\begin{equation}
\frac{dN}{dL_\gamma}=\rho_0
\Phi_0\left[\left(\frac{L_\gamma}{L_{\gamma \rm
b}}\right)^{\alpha_1} +\left(\frac{L_\gamma}{L_{\gamma \rm
b}}\right)^{\alpha_2}\right]^{-1},
\end{equation}
where $\rho_0=1.12\rm Gpc^{-3}yr^{-1}$ is the local  rate of GRBs,
and $\Phi_0$ is a normalization constant to assure the integral over
the luminosity function being equal to the local rate $\rho_0$. This
luminosity function breaks at $L_{\gamma \rm b}=2.25\times 10^{52}
\rm erg\,s^{-1}$, with indices $\alpha_1=0.65$ and $\alpha_2=2.3$
below and above the break. The rate distribution with redshift used
by L07 in obtaining this luminosity function is
 $S(z)=23 {e^{3.4z}}/({e^{3.4z}+22.0})$\citep{Porciani01}.}}
and \citet[][hereafter, G07]{Guetta07}{\footnote{The luminosity
function obtained by G07 has the same form as that of W10, but has
different parameter values, i.e., $\rho_0=0.27\rm Gpc^{-3}yr^{-1}$,
$a_1=-1.1$, $a_2=-3.0$ and $L_{*}=2.3\times 10^{51}\rm erg\,s^{-1}$.
This luminosity function is obtained based on the assumption that
the rate of GRBs follows the star formation history given by
\citet{RR99}, i.e.
\begin{equation}
S(z)= \left\{
\begin{array}{ll}
10^{0.75z} &z<1,\\
10^{0.75}  &z\geq 1.
\end{array}
\right.
\end{equation}}}. We find that the
GRB sample generated using the above approach with the luminosity
function and rate  redshift distribution given in L07 contain much
more triggered GRBs than the observed ones, while the G07 luminosity
function in combination with the relevant redshift distribution
gives rise to too few triggered GRBs. In other words, to maintain
the rate of triggered GRBs  of 250/yr by Fermi/GBM, $\rho_0$ is
required to be $\sim 0.2\, \rm Gpc^{-3}yr^{-1}$ and $\sim 5\, \rm
Gpc^{-3}yr^{-1}$ for L07 and G07 respectively, which are beyond the
$3\sigma$ error of the best--fit values ($1.12\, \rm
Gpc^{-3}yr^{-1}$ and $0.27\, \rm Gpc^{-3}yr^{-1}$ for L07 and G07
correspondingly). So here we adopt the luminosity function and
redshift distribution of W10 in generating the GRB sample in our
simulation.

\section{Diffuse neutrino emission from our GRB sample}
\subsection{ Analytical calculation of neutrino flux}
To illustrate the dependence of neutrino flux on the  parameters, we
first present an  analytical approximation of calculating the
neutrino flux.

It has been suggested that protons can be accelerated up to
$\epsilon_p^{\rm ob} \sim 10^{20}$eV through internal or external
shocks \citep[e.g.][]{Waxman95, Vietri95, Wang08, Murase08}. The
accelerated protons collide with the fireball photons in the same
region and produce charged pions. The charged pions decay into four
final-state leptons via the processes $\pi^{+}(\pi^-)\rightarrow
\nu_{\mu}(\bar{\nu}_{\mu})+\mu^+(\mu^-)\rightarrow
\nu_{\mu}(\bar{\nu}_{\mu})+e^+(e^-)+\nu_e(\bar{\nu}_e)+\bar{\nu}_{\mu}(\nu_{\mu})$.
In the fireball comoving frame, the fractional energy loss rate of a
proton with Lorentz factor $\gamma_p$  via photopion production is
\begin{equation}
t_{p\gamma}^{-1}=\frac{c}{2\gamma_p^2}\int_{\tilde{E}_0}^\infty
d\tilde{E}\sigma_\pi
(\tilde{E})\xi(\tilde{E})\tilde{E}\int_{\tilde{E}/2\gamma_p}^{\infty}dxx^{-2}n(x),
\end{equation}
where $c$ is the speed of light, $\tilde{E}$ is the photon energy in
the rest frame of the  proton and $\tilde{E}_0$ is the threshold
energy of photonpion production. $\sigma_\pi$ is the cross section
of photopion production and $\xi$ is the inelasticity. $x$ and
$n(x)$ are the photon energy and the photon spectrum in the GRB
comoving frame respectively. When only the $\Delta$ resonance
channel for the photopion production is considered, the fraction of
the proton energy converted into the pion can be approximated by
\begin{equation}\label{fpg}
f_{p\gamma}(\epsilon_p^{\rm ob})\propto \frac{L_{\rm iso}}{E_{\rm peak}^{\rm ob}\Gamma^2 R}\times\left\{
\begin{array}{ll}
\left(\frac{\epsilon_p^{\rm ob}}{\epsilon_{pb}^{\rm ob}}\right)^{-\beta-1}, & \epsilon_p^{\rm ob} < \epsilon_{pb}^{\rm ob}\\
\left(\frac{\epsilon_p^{\rm ob}}{\epsilon_{pb}^{\rm ob}}\right)^{-\alpha-1},  & \epsilon_p^{\rm ob} \geq \epsilon_{pb}^{\rm ob}
\end{array}
\right.
\end{equation}
where $\Gamma$ is the bulk Lorentz factor of GRB and $R$ is the
internal shock radius which relates with the bulk Lorentz factor and
variability time $\delta t$ through $R=2\Gamma^2c\delta t$.  The
break energy $\epsilon_{pb}^{\rm ob}$ in the proton spectrum is
caused by the break in the photon spectrum. 
Correspondingly, a break occurs in the neutrino spectrum at energy
\begin{equation}
\epsilon_{\nu b}^{\rm ob}=7.5\times 10^{14}(1+z)^{-2}\Gamma_{2.5}^{2}E_{\rm peak, MeV}^{\rm ob,-1} \, \rm eV.
\end{equation}
Additionally, there is another break in the neutrino spectrum caused
by  the synchrotron cooling of charged pions in the magnetic field
of the dissipative fireball, which can be written as
\begin{equation}
\epsilon_{\nu c}^{\rm ob}=3.3\times 10^{17}(1+z)^{-1}L^{-1/2}\Gamma_{2.5}^2R_{14}\, \rm eV.
\end{equation}
The final neutrino spectrum (single flavor) of a GRB is
approximately a three-section power law with two breaks at
$\epsilon_{\nu b}^{\rm ob}$ and $\epsilon_{\nu c}^{\rm ob}$
respectively, i.e.,
\begin{equation}{\label{flu}}
\begin{split}
(\epsilon_{\nu}^{\rm ob})^2 \frac{dn_{\nu}}{d\epsilon_\nu^{\rm
ob}}\simeq & \frac{1}{8}{\frac{\eta_p F_{\gamma,\rm bol}}{\rm
ln(\epsilon_{p,max}/\epsilon_{p,min})}}
f_{p\gamma}(\epsilon_{\nu}^{\rm ob})\\
& \times \left \{
\begin{array}{lll}
(\epsilon_{\nu}^{\rm ob})^{-\beta-1}, & \epsilon_{\nu}^{\rm ob} < \epsilon_{\nu b}^{\rm ob}\\
(\epsilon_{\nu}^{\rm ob})^{-\alpha-1}, & \epsilon_{\nu b}^{\rm ob} < \epsilon_{\nu}^{\rm ob} < \epsilon_{\nu c}^{\rm ob}\\
(\epsilon_{\nu}^{\rm ob})^{-\alpha-3}, & \epsilon_{\nu c}^{\rm ob} <
\epsilon_{\nu}^{\rm ob}.
\end{array}
\right.
\end{split}
\end{equation}
where $\eta_p$ is the ratio between the energy in the accelerated
protons and the radiation energy, $\epsilon_{p,\rm max}$ and
$\epsilon_{p,\rm min}$ are respectively the maximum and minimum energy
of accelerated protons, and $F_{\gamma,\rm bol}$ is the bolometric
(1\,keV--10\,MeV) gamma-ray fluence of the GRB. The factor 1/8 comes
from the fact that half of the produced pions are charged pions and
each neutrino takes about 1/4 of the pion's energy.

\subsection{Numerical results}
In our numerical calculation, besides  the $\Delta$ resonance
channel, we also consider the contribution from the direct pion
channel and the multi pion channel. The cross--section for photopion
production is taken from \citet{Mucke00} and the branching ratio of
each channel is approximated as the same in H12. Muon cooling, pion
cooling and neutrino oscillation effects are also included in the
calculation. We calculate the GRB neutrino flux within the standard
internal shock model, i.e., we use $R=2\Gamma^2c\delta t$, similar
to the treatment by the IceCube Collaboration
\citep{IC10,IC11a,IC12}. We consider two cases of the choice of the
bulk Lorentz factor $\Gamma$, one is the relation $\Gamma=29.8E_{\rm
iso,52}^{0.51}$ that is obtained by \citet{Ghirlanda12}, and another
is the benchmark choice, i.e. $\Gamma=10^{2.5}$ for all GRBs. As in
H12, we fix $\delta t^{\rm ob}=\delta t(1+z)$ at 0.01s.  We assume
the accelerated proton spectrum to be a power law with an index of
$-2$. The baryon ratio is taken to be $\eta_p=10$.

We  calculate the neutrino fluence from each GRB in the generated
sample and the aggregated neutrino fluence from all the triggered
GRBs in the sample  is shown in Fig.3. To compare with the IceCube
limit from the combined analysis of IC40 and IC59 for 215 GRBs, we
have multiplied the aggregated fluence by a factor of 215/$N_{\rm
tri}$, where $N_{\rm tri}=250$ is the number of triggered GRBs in
our sample. The aggregated neutrino fluence peaks around 1\,PeV and
reach a level of $\sim 0.1$GeVcm$^{-2}$ for both choices of
$\Gamma$. For comparison, we also show the fluence limit of IC40+59
in Fig.3 (dashed lines). The 90\% upper limit on the fluence is
obtained  by requiring that the upgoing muon neutrino event
number{\footnote{The upgoing muon neutrino number is obtained by
summing up the contribution from every individual GRBs in the energy
range of 0.1-3 PeV. For these upgoing neutrinos, we take
$90^{\circ}-180^{\circ}$ angle-averaged effective area for IC40 and
IC59 in the calculation. The ratio of the number of GRBs calculated
with the effective area of IC40 and  IC59 is set to 117:98, in
alignment with the realistic number ratio of GRBs that occurred
during IC40 and IC59 operation period. }} observed by IC40+59 should
be smaller than $N_{\rm limit}=1.9$.  The non-detection of
neutrinos by IC40+59 implies $\eta_p\la20$. We note that the
aggregated neutrino fluence of 215 GRBs in our former work H12 (see
Fig.~2 in H12) is a factor of two lower and peaks at higher energy.
The difference is due to that, in that work the luminosity and bulk
Lorentz factor  of all the GRBs without measured redshift are fixed
as $10^{52}$ergs$^{-1}$ and $10^{2.5}$ respectively.

The diffuse neutrino flux can be obtained by multiplying the
aggregated neutrino fluence of the all--sky GRBs occurred in one
year by a factor of $1/4\pi\,\rm yr\simeq 2.5\times 10^{-9} \rm
s^{-1}sr^{-1}$. Fig.~4 presents the diffuse neutrino flux from all
GRBs in our generated sample (solid lines),  which consists of the
flux contributed  by the triggered GRBs (dashed line) and the flux
contributed by untriggered GRBs (dash-dotted line). The top panel
corresponds to the case that assumes $\Gamma=29.8E_{\rm
iso,52}^{0.51}$ \citep{Ghirlanda12}  and the bottom to the case of
$\Gamma=10^{2.5}$. In both cases, the total flux around PeV energy
are $\sim \rm 10^{-9} GeVcm^{-2}s^{-1}sr^{-1}$. The flux produced by
the untriggered GRBs is about a factor of two higher than that of
the triggered ones.

{In Fig.~5, we show  the contribution to the diffuse neutrino flux
by untriggered  GRBs with different luminosity separately. As can be
seen, GRBs in the intermediate luminosity range
($10^{51}-10^{53}{\rm erg s^{-1}}$) contribute the largest fraction
of the diffuse neutrino flux in both cases of the choice of the bulk
Lorentz factor. These GRBs do not trigger the detectors either due
to occurring at relatively high redshifts, and/or due to having high
spectral peaks (e.g., $E_{\rm peak}^{\rm ob}\gg 300\,\rm KeV$).
Besides, the occulted bright GRBs also contribute a significant
fraction to the total neutrino flux. }

\subsection{Comparison with two PeV neutrinos}
The IceCube Collaboration recently reported the detection of two
cascade neutrinos  in the range 1--10 PeV\citep{Aya12}, which are
likely to be electron (or anti-electron) neutrinos. Several
scenarios have been proposed to explain the origin of these two
neutrinos \citep{Barger12, Baerwald12b, Bhattacharya12,
Roulet12,Cholis12}. Since  the energies of the reported two
neutrinos are close to the peak energy of the typical GRB neutrino
spectrum,  we here study the possibility that they originate from
diffuse GRB neutrinos. In studying the possible GRB origin, one must
assure that the stacked neutrino flux of those 215 triggered GRBs
observed by IceCube do not exceed the IC40+59 limit. The expected
number of diffuse neutrinos from GRBs in the energy range 1--10PeV
can be obtained by
\begin{equation}\label{Ndiff}
N_{\rm diff}=\int_{1\rm PeV}^{10\rm
PeV}S(\epsilon_{\nu_e})\left(\frac{dn_{\nu_e}}{d\epsilon_{\nu_e}}\right)_{\rm
diff}d\epsilon_{\nu_e},
\end{equation}
where $\left(\frac{dn_{\nu_e}}{d\epsilon_{\nu_e}}\right)_{\rm diff}$
is the  number spectrum of diffuse electron neutrinos (in unit of
$\rm cm^{-2}s^{-1}sr^{-1}eV^{-1}$) and $S(\epsilon_{\nu_e})$ is the
IceCube exposure of electron neutrinos at ultra-high energies in 2
years (2010--2012)\citep{Aya12}, which  includes the contribution by
the Glashow resonance. Notably, since the analysis approach for
diffuse cascade neutrinos is different from that for upgoing
neutrinos from triggered GRBs, the IceCube  effective area is
significantly different. Based on the total diffuse neutrino flux we
obtained above, we get $N_{\rm diff}\approx 0.1$ for the case of
$\Gamma=29.8E_{\rm iso, 52}^{0.51}$ and $N_{\rm diff}\approx 0.2$
for the case of $\Gamma=10^{2.5}$ ($\eta_p=10$ is assumed). Since
the non--detection of GRBs by IC40+59  implies $\eta_p < 20$ (see \S
3.2), we have $N_{\rm diff}<0.2$ or $0.4$ respectively, which is
insufficient to account for the reported two neutrinos. There is
still a small possibility that these two neutrinos arise from, for
instance, some very strong GRBs that happen to be occulted by the
Earth and hence do not trigger the detectors. In addition, strong
statistical fluctuation can also possibly lead to detection of one
or two events.

\subsection{Effect of variation in luminosity function and  redshift distribution}

There is significant variation in the luminosity function $\phi (L)$
and the rate distribution with redshift $\rho(z)$ in the literature(e.g.
\citealt{Firmani04}, G07, L07, W10,  \citealt{Zitouni08, Cao11}).
The variation may be caused by different approaches in obtaining the
luminosity function and different spectral properties as well as
trigger thresholds being used in the calculation.

Under the requirement that the total neutrino fluence from the
triggered GRBs do not exceed the upper limit placed by IceCube, the
possibility that GRBs are responsible for the two events increases
if the fraction of neutrino flux from the untriggered GRBs
increases. This requires more dim and/or distant GRBs to take place.
Therefore, luminosity functions with steep slopes or rate
distribution with higher rate at high redshift  are favorable for
producing more dim, untriggered GRBs. As  the slopes in the
luminosity function of L07 below and above the break luminosity
$L_b$ are shallower  than that in W10, adopting this luminosity
function will lead to a smaller fraction of untriggered GRBs.
However, the rate redshift distribution used in L07 suggests higher
rate of GRBs at high redshift, which, on the contrary,  will lead to
a higher fraction of untriggered GRBs. We estimate that the neutrino
flux produced by untriggered GRBs is about 90\% of that of triggered
GRBs for the luminosity function and rate distribution in L07. For
the luminosity function and rate distribution in G07, this factor
increases to 2.5 because of the steep slope in the luminosity
function of G07 and higher GRB rate at high redshift. Even
considering the variation in the luminosity function and rate
distribution in the literature, the diffuse neutrino flux from GRBs
may be still insufficient to account for the two neutrinos.

\section{Discussions}
Other possible sources of dim, untriggered GRBs are low--luminosity
GRBs (LLGRBs) and Population III (Pop III) GRBs. LLGRBs are GRBs
with luminosity $\lesssim 10^{49}$ergs$^{-1}$, which constitute a
distinct population from the normal GRBs\citep[e.g.][]{Soderberg07,
Liang07,Dai09,Virgili09}. Although only several cases are observed
within $\sim 200$Mpc so far, which is mainly due to their low
luminosity, the intrinsic local event rate of LLGRBs is suggested to
be about two orders of magnitude higher than that of the normal
GRBs\citep[e.g.][]{Liang07,Dai09,Virgili09}. Thus, the large local
population could compensate for the low energy in LLGRBs, and  lead
to a significant amount of neutrino flux \citep[e.g.][]{Murase06b,
Gupta07,Murase08,Liu11}. In \citet{Liu11}, we calculated the diffuse
neutrino flux from LLGRBs for different LLGRB luminosity functions
and found that  the flux can reach a level of $\sim 10^{-9}\rm
GeVcm^{-2}s^{-1}sr^{-1}$ or higher around PeV energies, depending on
the parameters of LLGRBs that are used, such as the choice of
$\Gamma$. We also found that  most LLGRBs are untriggered and the
energy budget of gamma rays in untriggered LLGRBs can be several
times  larger than that in the triggered ones. Thus more than
0.2--0.4 electron neutrinos in the range 1--10\,PeV could come from
LLGRBs in principle. However, since the origin of LLGRBs are not
clear \citep[e.g.][]{Wang07} and the properties of LLGRBs have large
uncertainties, we should be cautious about this estimate.

Pop III GRBs are thought to arise from the death of the first stars
that locate at high redshift ($z\gtrsim 10$). It is suggested that
the their luminosity  is typically a few times $10^{52}\,\rm
ergs^{-1}$  and their duration is  $\gtrsim 10^4$s\citep{Meszaros10,
Komissarov10, Suwa11},  so they may release energy up to
$10^{56-57}$erg in one burst. At such large distances, their
gamma-ray flux is below $\sim 10^{-8}\,\rm ergcm^{-2}s^{-1}$, so
they may not be able to trigger current GRB detectors. Thus, if Pop
III GRBs can produce high energy neutrinos, these neutrinos will
contribute to the diffuse background.  It has been shown by
\citet{Iocco08, Gao11} that the neutrino flux in 1-10 PeV range can reach
$\sim 10^{-9}\,\rm GeV cm^{-2}s^{-1}sr^{-1}$, or even $\sim 10^{-8}\,\rm
GeV cm^{-2}s^{-1}sr^{-1}$ when some preferred
parameters are used, e.g., with large progenitor mass and/or high
interstellar medium density. However,  since the flux peaks at $\sim
10^{17}$ eV, one would expect even stronger flux at $\sim 10^{17}$
eV, which is inconsistent with the non-detection by IceCube at such
energies. Moreover, the rate of Pop III GRBs is quite uncertain. The
Pop III GRB rate adopted in \citet{Gao11} is from \citet{Bromm06},
which obtains a much higher rate  than that in \citet{deSouza11} and
\citet{Campisi11}. So whether Pop III GRBs are responsible for the
two neutrinos remains to be studied.

Finally we note that we only consider the neutrino emission under
the standard internal shock model. For other models for the GRB
prompt emission, such as the photosphere emission models
\citep[e.g.][]{Rees05} or magnetically reconnection models
\citep[e.g.][]{Zhang&Yan11}, the efficiency for producing neutrinos
$f_{p\gamma}$ is different \citep[e.g.][]{Wang09, Zhang12}, due to
different dissipation radii $R$ involved. However, the ratio between
the flux from triggered GRBs and that from untriggered GRBs is not
expected to change. Therefore, the diffuse neutrino flux can not be
readily increased under the constraint that no neutrinos are
detected from from triggered GRBs.

\section{Conclusions}
In this paper, we studied the diffuse neutrino emission from GRBs in
light of the recent report that two PeV cascade events are detected
by IceCube. We first generate a GRB sample that reproduces the
realistic properties of the observed GRBs by Fermi/GBM. An
accompanying sample of untrigged GRBs is generated simultaneously.
Then we calculated the neutrino flux from the triggered and the
untriggered GRBs in the sample within the standard internal shock
scenario. The advantage of this approach is that the spectral
properties (i.e. the spectral peak and indices) of every individual
GRBs can be taken into account properly in the calculation. We found
that the neutrino flux from the untriggered GRBs is about a factor
of two larger than that of the triggered GRBs. Under the constraint
of the non-detection of any upgoing muon neutrinos from 215
triggered GRBs in the IC40+59 combined analysis, the total diffuse
neutrinos flux (single flavor) can reach a level of $\sim10^{-9}\rm
GeVcm^{-2}s^{-1}sr^{-1}$ at most. Such a flux can only give rise to
$\lesssim 0.2-0.4$  cascade  neutrino events in 1--10\,PeV for
IceCube 2-years exposure. Low--luminosity GRBs and/or Pop III GRBs
can also contribute to the diffuse neutrino flux, but large
uncertainties in their properties prevents us from drawing a firm
conclusion.

\acknowledgments We thank Kohta Murase, Jun Kakuwa,  Aya Ishihara,
Shigeru Yoshida and Ilias Cholis for useful discussions. This work
is supported by the 973 program under grant 2009CB824800, the NSFC
under grants 10973008, 11273016 and 11033002, the Excellent Youth
Foundation of Jiangsu Province and the Fok Ying Tung Education
Foundation.

\clearpage

\clearpage
\begin{figure}
\plotone{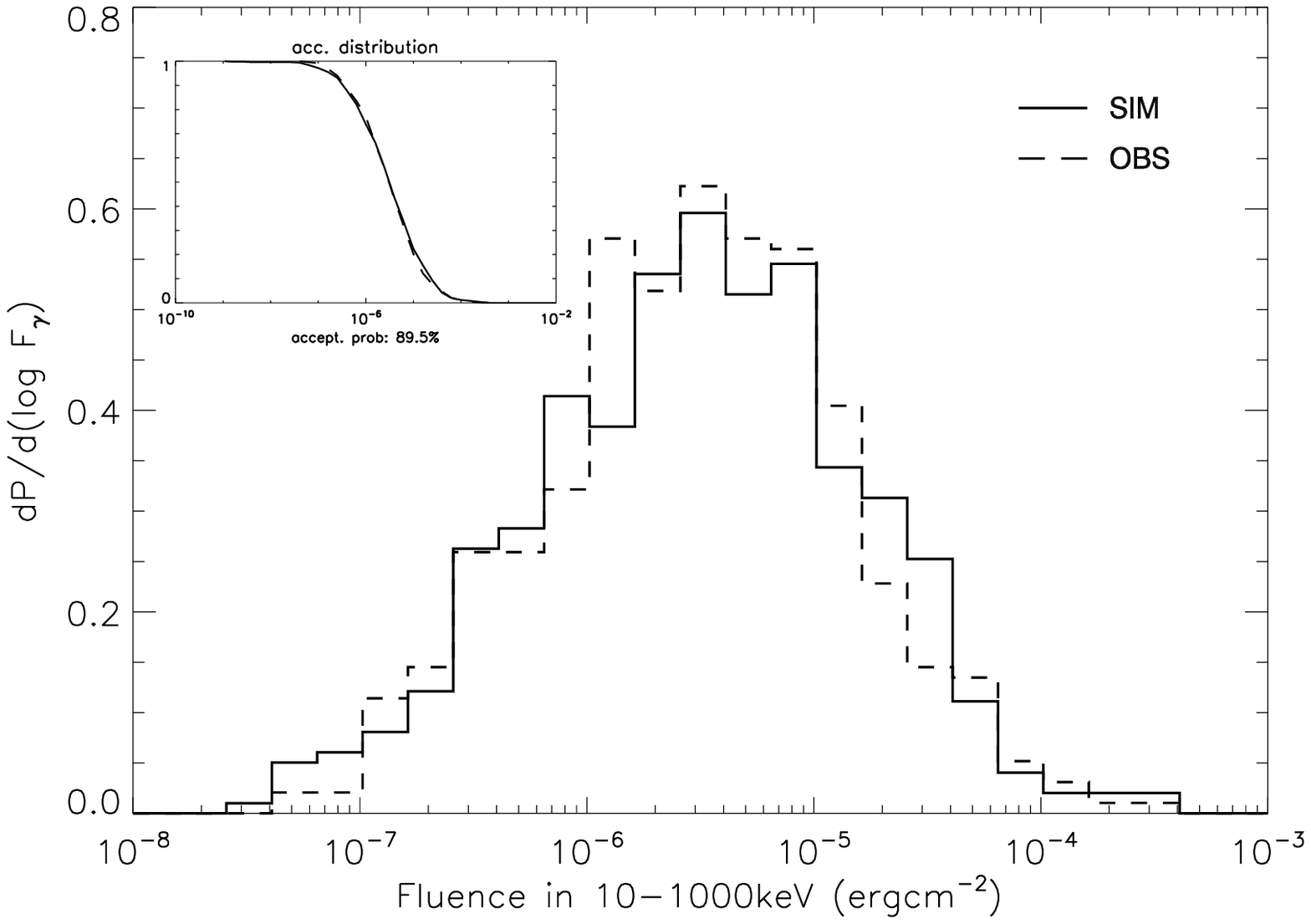} \caption{Comparison of the fluence (10-1000\,keV)
distribution of GRBs in the generated sample and in the observed
sample by Fermi/GBM. The solid line and dashed line represent the
triggered GRBs in our generated sample and the observed sample
respectively. The K-S test gives a  probability of 89.5\% for the
agreement the two distributions. The inset figure shows the
distributions of accumulated fluence of these two
samples.\label{fig1}}
\end{figure}

\begin{figure}
\plotone{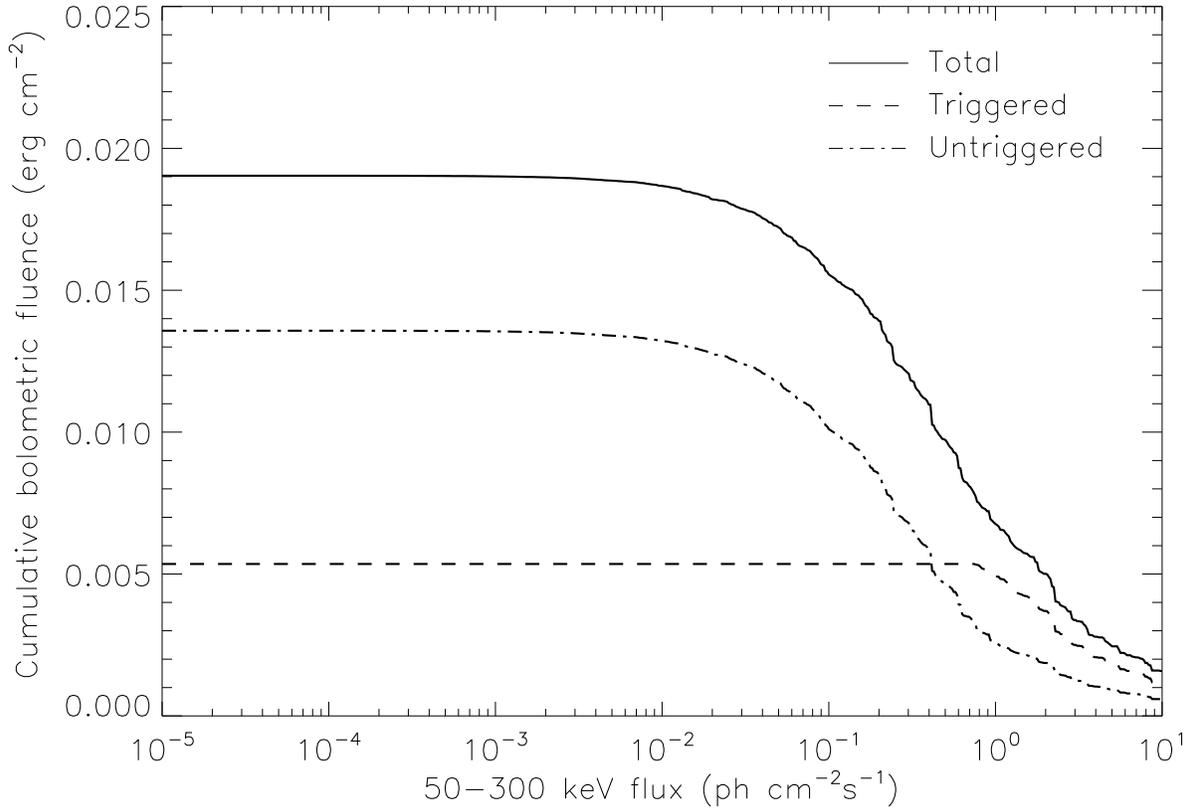} \caption{ Cumulative bolometric fluence
(1-10000\,keV) as a function of the flux  (50-300KeV) of GRBs in the
generated sample. The triggered GRBs, untriggered GRBs and the total
are represented by the dashed line,  dot--dashed line and  solid
line respectively. Since about 35\% GRBs with flux above the trigger
threshold are ascribed to untriggered GRBs due to occultation by the
Earth and the South Atlantic Anomaly  passages, the cumulative
fluence of untriggered GRBs is non--zero below the Fermi/GBM trigger
threshold $0.74\,\rm ph cm^{-2}s^{-1}$.\label{fig2}}
\end{figure}
\clearpage

\begin{figure}
\plotone{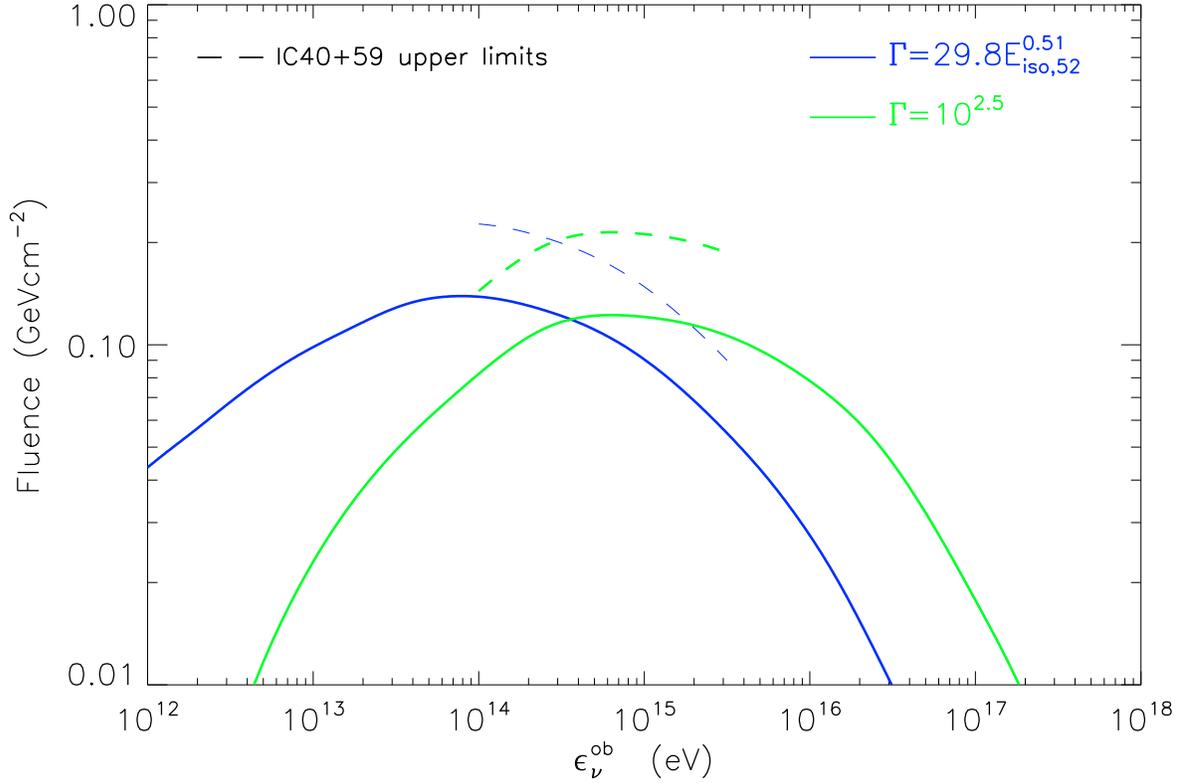} \caption{The aggregated neutrino fluence spectrum
(single flavor) from  triggered GRBs in our generated sample. The
blue and green solid lines represent the fluence obtained assuming
$\Gamma=29.8E_{\rm iso, 52}^{0.51}$ (Ghirlanda et al. 2012)  and the
benchmark value $\Gamma=10^{2.5}$ respectively. The corresponding
dashed lines represent the 90\% upper limits from  IC40+59. In the
calculation, we assumed that the proton spectrum is a power law with
index of $-2$ and $\eta_p=10$. \label{fig3}}
\end{figure}
\clearpage

\begin{figure}
\vskip -1cm \epsscale{0.8} \plotone{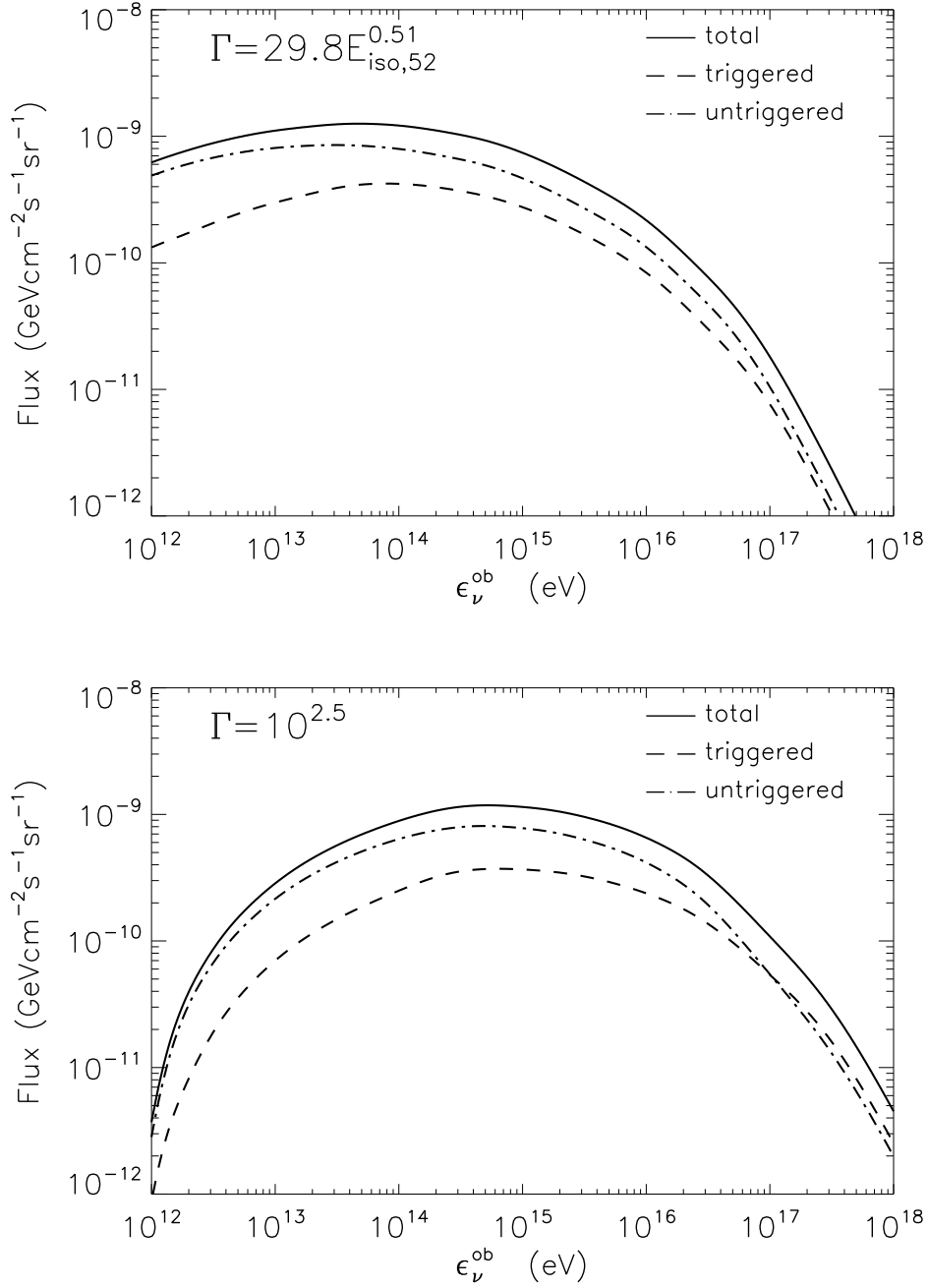} \caption{The diffuse neutrino flux
from GRBs in our generated sample. The dashed line and the
dash--dotted line represent the contribution by the triggered GRBs
and untriggered GRBs respectively, and the  solid line represnets
the total flux. In the top panel  $\Gamma=29.8E_{\rm iso,
52}^{0.51}$ is used, while in the bottom panel  $\Gamma$ is fixed at
$10^{2.5}$. We assumed that the proton spectrum is a power law with
index of $-2$ and $\eta_p=10$. \label{fig4}}
\end{figure}
\clearpage

\begin{figure}
\plotone{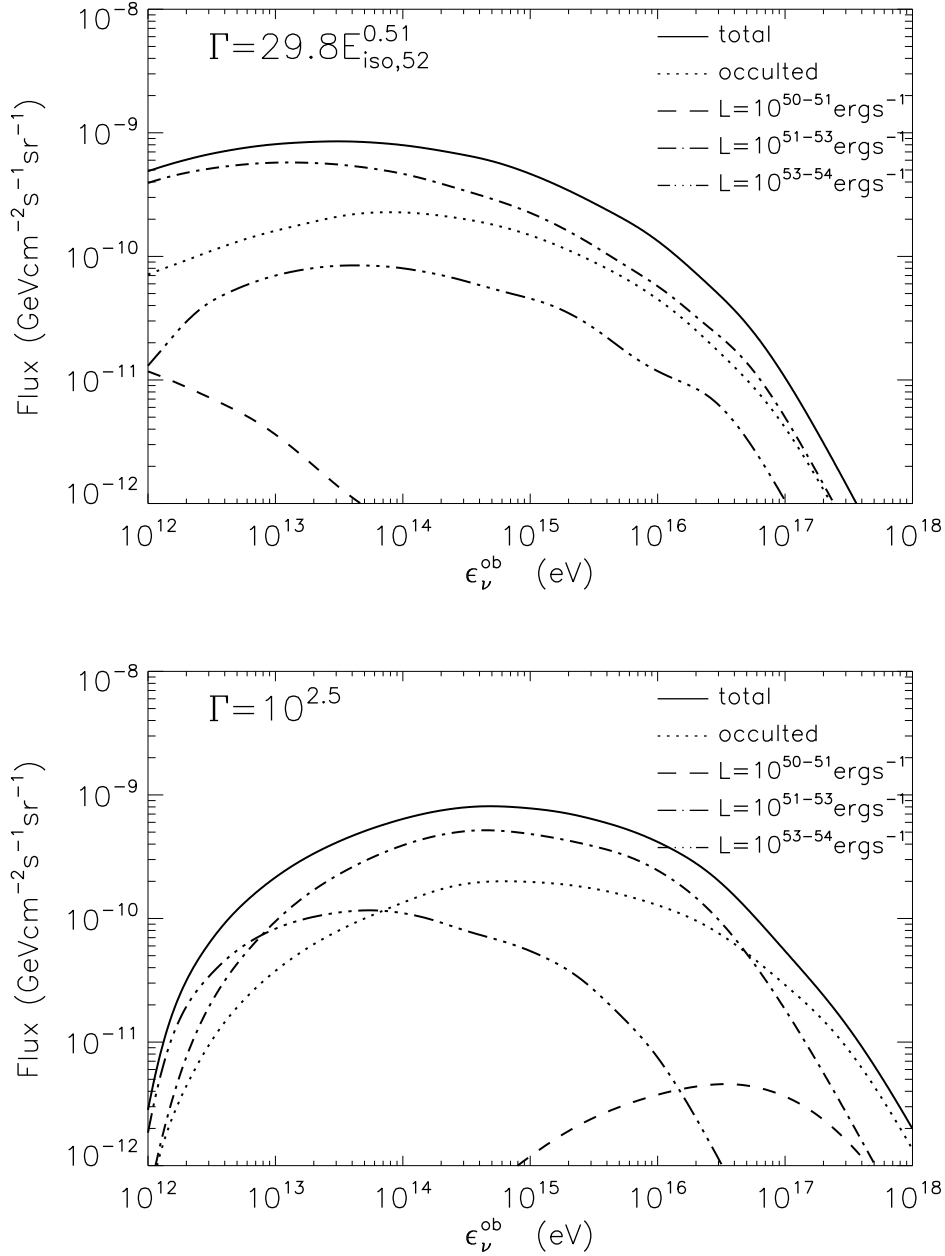} \epsscale{0.8} \caption{Contribution to the
diffuse neutrino flux by untriggered GRBs with different luminosity.
The dotted lines represent the contribution by the occulted bright
GRBs. The dashed lines, dash-dotted lines and the long dash-dotted
lines represent the contributions from GRBs in different luminosity
ranges respectively (not including those occulted GRBs). The solid
lines represent the sum of  the four components. ). \label{fig5}}
\end{figure}

\end{document}